\documentclass [12pt]{article}
\usepackage{dafne99pro}
\usepackage{latexsym}
\usepackage{hyperref}
\usepackage{tabularx}
\input boxedeps
\SetepsfEPSFSpecial 

\HideDisplacementBoxes
\newcommand{\dafne}{DA$\Phi$NE}
\newcommand{\CP}{\textsf{CP}}
\newcommand{\epsp}{\ensuremath{\varepsilon^{\prime}}}
\newcommand{\etal}{{\em et al.}}

\newcommand{\gevcc}{\hbox{ GeV}\!/\!c^2}
\newcommand{\gev}{\hbox{ GeV}}

\newcommand{\ev}{\hbox{ eV}}

\newcommand{\tev}{\hbox{ TeV}}

\newcommand{\ps}{\hbox{ ps}}

\newcommand{\cm}{\hbox{ cm}}
\newcommand{\mm}{\hbox{ mm}}

\newcommand{\fb}{\hbox{ fb}}

\newcommand{\m}{\hbox{ m}}

\newcommand{\lum}{\hbox{ cm}^{-2}\hbox{ s}^{-1}}
\newcommand{\eqn}[1]{(\ref{#1})}

\def\ltap{\mathop{\raisebox{-.4ex}{\rlap{$\sim$}} 
\raisebox{.4ex}{$<$}}}

\def\ket#1{| #1\rangle}
\newcommand{\real}[1]{\ensuremath \mathrm{Re}(#1)}

\newcommand{\cfrac}[2]{\textstyle \frac{#1}{#2}}

\newcommand{\hepex}[1]{hep-ex/#1}
\newcommand{\hepph}[1]{hep-ph/#1}

\newcommand{\heplat}[1]{hep-lat/#1}

\def\prl#1#2#3{\frenchspacing{\it Phys. Rev. Lett. }{\bf #1}, #2 (19#3)}

\def\pra#1#2#3{\frenchspacing{\it Phys. Rev. A}{\bf #1}, #2 (19#3)}
\def\prc#1#2#3{\frenchspacing{\it Phys. Rev. C}{\bf #1}, #2 (19#3)}
\def\pl#1#2#3{\frenchspacing{\it Phys. Lett. }{\bf #1}, #2 (19#3)}
\def\np#1#2#3{\frenchspacing{\it Nucl. Phys. }{\bf #1}, #2 (19#3)}

\def\arnps#1#2#3{\frenchspacing{\it Ann. Rev. Nucl. Part. Sci. }{\bf #1}, #2 (19#3)}
\def\ib#1#2#3{{\bf #1}, #2 (19#3)}

\begin{document}
\title{
\CP\ VIOLATION AND RARE DECAYS}

\author{Chris Quigg\thanks{\quad E-mail address: 
\textsf{quigg@fnal.gov}.\hfill\fbox{\textsf{FERMILAB--CONF--00--002--T}}} \\
{\em Fermi National Accelerator Laboratory} \\ {\em P.O. Box 500, Batavia, 
Illinois 60510 USA}}
\maketitle
\baselineskip=14.5pt
\begin{abstract}
After a brief essay on the current state of particle physics and 
possible approaches to the opportunities before us,
I summarize the contributions to the Third Workshop on Physics and 
Detectors for \dafne\ that deal with \CP\ Violation and Rare Decays.
\end{abstract}
\baselineskip=17pt
\section{Prologue}
This is my first visit to Frascati, and it has been thrilling for me
to be in this laboratory of such great significance to the development
of electron--positron colliding beams.  I have greatly enjoyed the
engagement between theory and experiment in the laboratory, the spirit
of this week's workshop, and the sense of anticipation for the
\dafne\ physics program.

I am delighted that our hosts have arranged a post-workshop visit to the 
Vatican Museums and the Sistine Chapel.  In February of 1988, I 
brought my children to Rome during their school vacation.  In the 
course of our 
visit to the Vatican Museums, we came across a temporary exhibition in 
which a panel displayed a passage that struck me as so 
remarkable and perceptive that I copied it down in my notebook:

\begin{center}
    {\footnotesize 
    \begin{tabularx}{\linewidth}{X|X}
    Il \emph{De Rerum Natura} di Lucrezio \`{e} la prima grande opera di 
    poesia in cui la conoscenza del mondo diventa dissoluzione della 
    compattezza del mondo, percezione di ci\`{o} che \`{e} infinamente muto e 
    mobile e leggero.
    
    Lucrezio vuole scrivere il poema della materia ma ci avverte subito 
    che la vera realt\`{a} di questa materia \`{e} fatta di corpuscoli invisibili.  
    \`{E} il poeta della concretezza fisica, vista nella sua sostanza 
    permanente e immutibile, ma per prima cosa ci dice che il vuoto \`{e} 
    altrettanto concreto che i corpi solidi.
    
    La pi\`{u} grande preoccupazione di Lucrezio sembra quella di evitare che 
    il peso della materia ci schiacci.
        \begin{flushright}
            Italo Calvino, ``Leggerezza,'' in\\
            \emph{Lezioni Americane} (1988)\\[-12pt]
        \end{flushright}
    &
    The \emph{De Rerum Natura} of Lucretius is the first great work of 
    poetry in which knowledge of the world tends to dissolve the solidity 
    of the world, leading to a perception of all that is infinitely 
    minute, light, and mobile.  
    
    Lucretius set out to write the poem of physical matter, but he warns 
    us at the outset that this matter is made up of invisible particles.  
    He is the poet of physical concreteness, viewed in its permanent and 
    immutable substance, but the first thing he tells us is that emptiness 
    is just as concrete as solid bodies.
    
    Lucretius' chief concern is to prevent the weight of matter from 
    crushing us \ldots 
        \begin{flushright}
             ~\phantom{Calvino}~ \\ Italo Calvino, ``Lightness,'' in\\
             \emph{Six Memos for the Next Millennium}\\[-12pt]
        \end{flushright}
    \end{tabularx}
    }
\end{center}
\vspace*{6pt}
The text is taken from the Charles Eliot Norton Lectures that Calvino 
prepared to deliver at Harvard.\footnote{He died on September 19, 
1985, just before departing for Cambridge.} Back in Berkeley, I found 
a copy of the American translation, \textit{Six Memos for the Next 
Millennium,} one of the great short works of literary analysis and 
appreciation.\cite{italo}  In the translation, I found that the passage on a panel 
in the Vatican Museums concludes, ``The poetry of the invisible, of 
infinite unexpected possibilities---even the poetry of 
nothingness---issues from a poet who had no doubts whatever about the 
physical reality of the world.'' For us who have no doubts whatever 
about the physical reality of the world, who experience daily the 
poetry of the invisible, of infinite unexpected possibilities---even 
of nothingness---it is inspiring to see our passion and our confidence 
in Nature reflected in Calvino's scan of Lucretius.  I hope that each 
of you will have a similar epiphany in the Vatican Museums tomorrow.

\section{What Strange Particles Have Been Telling Us \ldots \\
(When We Have Known How to Listen)}
For half a century, the strange particles have offered us important 
clues to the nature of matter and the character of the forces that 
shape the world.  It is interesting to consider some of the phenomena 
that pointed the way to our current picture of constituents and 
interactions, and the lessons we drew---or could have drawn---from 
them.\newpage
\begin{itemize}
    \item  \textit{Strangeness.}
 
    At a time when the significance of isospin was incompletely 
    appreciated, the $K$ mesons and hyperons showed us the importance of an 
    imperfectly conserved quantum number, and of the notion that 
    different interactions can respect different symmetries.
    
    As our understanding of internal symmetries matured, strange particles 
    pointed the way to $SU(3)_{\mathrm{flavor}}$ as a classification 
    symmetry for the hadrons.  The observation in the 1960s that all mesons fit 
    into $SU(3)_{\mathrm{flavor}}$ singlets and octets, and that all 
    baryons fit into $SU(3)_{\mathrm{flavor}}$ singlets, octets, and 
    decimets, pointed to the quark model, with its rule that mesons 
    are composed of $q\bar{q}$ and baryons of $qqq$.  It is important 
    to note that the quark model gives us a new view of additive 
    quantum numbers, not as quantities carried by particles---to be 
    loaded and unloaded---but as defining attributes of the 
    fundamental constituents, the quarks.
    
    The conflict between the symmetric $sss$ wave function of the 
    $\Omega^{-}$ and the Pauli principle led to the introduction of 
    color as a hitherto unobserved quantum number and, eventually, to 
    the development of quantum chromodynamics as the theory of the 
    strong interactions.

    \item  \textit{Kaon Decays}
    
    The $\tau/\theta$ puzzle challenged the implicit notion that the 
    fundamental interactions are invariant under parity, and set in 
    motion the chain of investigations that gave us the $V-A$ theory 
    of the (charged-current) weak interactions.
    
    The identification of $K_{S}$ and $K_{L}$ provided the 
    opportunity to observe quantum-mechanical superposition effects 
    over macroscopic distances and focused attention on \textsf{CP} 
    eigenstates.  
    
    The very different rates for the decays $K_{S} \to 
    \pi^{+}\pi^{-}$ and $K^{+} \to \pi^{+}\pi^{0}$ gave evidence for 
    the $\Delta I = \cfrac{1}{2}$ rule in nonleptonic weak decays.  Though we 
     saw it only in retrospect, the nonleptonic ``octet'' 
    enhancement that accounts for the $\Delta I = \cfrac{1}{2}$ rule requires 
    colored quarks and the short-distance effects of the strong 
    interaction.
    
    The notion of Cabibbo universality gave rise to the idea of 
    mixing between quark generations and laid the foundation for a 
    theory of weak interactions based (in part) on $SU(2)_{L}$ 
    symmetry, which gives the framework for a connection between quarks 
    and leptons.
    
    \item  \textit{Rare Decays}
    
    The suppression of the strong decay $\phi \to \pi^{+}\pi^{-}$ gave 
    us clues about the character of the strong interaction, embodied 
    in the Okubo--Zweig--Iizuka rule, and offered a template for 
    understanding the inhibited decays of $J/\psi$ and $\Upsilon$ into 
    hadrons.
    
    The absence of flavor-changing neutral currents inspired the 
    second coming of charm and the Glashow--Iliopoulos--Maiani 
    mechanism.
    
    The agreement between the observed rate for the decay $K_{L} \to 
    \mu^{+}\mu^{-}$ and conventional expectations indicated that the 
    charmed quark could not be arbitrarily heavy, and permitted an 
    estimate $m_{c} \approx 1.5\gevcc$. 

    \item  \textsf{CP} \textit{Violation}
    
    The observation of the \CP-violating decay $K_{L} \to \pi^{+}\pi^{-}$ 
    suggested the need for at least three doublets of quarks 
    (therefore of leptons) and gave us reason to explore scenarios for 
    cosmic baryogenesis.
    
    The discovery that the direct \CP-violating parameter $\epsp \neq 
    0$ demonstrates that a superweak interaction is not the (only) 
    source of \CP\ violation, and supports the phase in the 
    quark-mixing matrix as the dominant origin of \CP\ violation.

    \item  $K^{0}\hbox{--}\bar{K}^{0}$ \textit{Mixing}
    
    The degree of $K^{0}\hbox{--}\bar{K}^{0}$ mixing, which is to say 
    the $K_{L}\hbox{--}K_{S}$ mass difference, together with the value 
    of the \CP-violating parameter $\varepsilon$, offered a hint that 
    the mass of the top quark was quite large.
\end{itemize}

Not all of these items represent finished business.  What clues have 
we missed?  What is to come?

\section{Our Picture of Matter}
Twenty-five years after the November Revolution, our understanding of
physical phenomena is grounded in the identification of fundamental 
constituents, at the current limits of resolution of about 
$10^{-18}\m$, and a few fundamental forces.  The constituents---our 
elementary particles---are the pointlike quarks
	\begin{equation}
		\left(
		\begin{array}{c}
			u  \\
			d
		\end{array}
		 \right)_{L} \;\;\;\;\;\;
		\left(
		\begin{array}{c}
			c  \\
			s
		\end{array}
		 \right)_{L} \;\;\;\;\;\;
		\left(
		\begin{array}{c}
			t  \\
			b
		\end{array}
		 \right)_{L}		 
	\end{equation}
and leptons
		\begin{equation}
		\left(
		\begin{array}{c}
			\nu_{e}  \\
			e^{-}
		\end{array}
		 \right)_{L} \;\;\;\;\;\;
		\left(
		\begin{array}{c}
			\nu_{\mu}  \\
			\mu^{-}
		\end{array}
		 \right)_{L} \;\;\;\;\;\;
		\left(
		\begin{array}{c}
			\nu_{\tau}  \\
			\tau^{-}
		\end{array}
		 \right)_{L}		 
	\end{equation}
with strong, weak, and electromagnetic interactions specified by
$SU(3)_{c}\otimes SU(2)_{L}\otimes U(1)_{Y}$ gauge symmetries.

It is instructive to consider the agenda of particle physics today 
under four rubrics:
\begin{itemize}
    \item  \textit{Elementarity.} Are the quarks and leptons 
    structureless, or will we find that they are composite particles 
    with internal structures that help us understand the properties 
    of the individual quarks and leptons?

    \item  \textit{Symmetry.} One of the most powerful lessons of the 
    modern synthesis of particle physics is that (local) symmetries prescribe 
    interactions.  Our investigation of symmetry must address the 
    question of which gauge symmetries exist (and, eventually, why).  
    We must also understand how the symmetries are hidden from us in 
    the world we inhabit.  For the moment, the most urgent problem in 
    particle physics is to complete our understanding of electroweak 
    symmetry breaking by exploring the 1-TeV scale.  This is the 
    business of the experiments at LEP2, the Tevatron Collider, and 
    the Large Hadron Collider.

    \item  \textit{Unity.} In the sense of developing explanations 
    that apply not to one individual phenomenon in isolation, but to 
    many phenomena in common, unity is central to all of physics, and 
    indeed to all of science.  At this moment in the development of 
    particle physics, the quest for unity takes several forms.  
    First, we have the fascinating possibility of gauge coupling 
    unification, the idea that all the interactions we encounter have 
    a common origin and thus a common strength at suitably high 
    energy.  Second, there is the imperative of anomaly freedom in 
    the electroweak theory, which urges us to treat quarks and 
    leptons together, not as completely independent species.  Both of 
    these ideas are embodied, of course, in unified theories of the 
    strong, weak, and electromagnetic interactions, which imply the 
    existence of still other forces---to complete the grander gauge 
    group of the unified theory---including interactions that change 
    quarks into leptons.  The third aspect of unity is the idea that 
    the traditional distinction between force particles and 
    constituents might give way to a unified understanding of all the 
    particles.  The gluons of QCD carry color charge, so we can 
    imagine quarkless hadronic matter in the form of glueballs.  
    Beyond that breaking down of the wall between messengers and 
    constituents, supersymmetry relates fermions and bosons.  Finally, 
    we desire a reconciliation between the powerful outsider, gravity, 
    and the forces that prevail in the quantum world of our everyday 
    laboratory experience.

    \item \textit{Identity.} We do not understand the physics that sets
    quark masses and mixings.  Although we are testing the idea that the
    phase in the quark-mixing matrix lies behind the observed \CP\ 
    violation, we do not know what determines that phase.  The 
    accumulating evidence for neutrino oscillations presents us with 
    a new embodiment of these puzzles in the lepton sector.  At 
    bottom, the question of identity is very simple to state: What 
    makes an electron and electron, and a top quark a top quark?
\end{itemize}

One aspect of the problem of identity is the origin of mass.  In 
particle physics, we know the challenge of explaining many different 
kinds of mass.  The masses of the hadrons are (in principle, and with 
increasing precision in practice) understood from QCD in terms of the 
energy stored to confine a color-singlet configuration of quarks in a 
small volume.\cite{fwpt}  We also have an excellent understanding of 
the masses of the electroweak gauge bosons $W^{\pm}$ and $Z^{0}$ as 
consequences of electroweak symmetry breaking, in terms of a single 
weak mixing parameter, $\sin^{2}\theta_{W}$.\footnote{Although for the
moment we take this parameter from experiment, we understand how it
arises in a unified theory.  Indeed, in a unified theory we can hope
to understand the parameter $\Lambda_{\mathrm{QCD}}$ that sets the
scale of the hadron masses.} When we get to the question of quark and
(charged) lepton masses, however, our understanding is considerably
more primitive.  For each of these, we require not just the scale of
electroweak symmetry breaking, but a distinct and apparently arbitrary
Higgs-fermion-antifermion Yukawa coupling to reproduce the fermion
mass.  For neutrinos, which may be their own antiparticles, there are
still more possibilities for new physics to enter, and, for the
moment, more room for bafflement.

\section{The Problems of Mass, and of Mass Scales}
As we have just remarked, electroweak symmetry breaking sets the 
values of the $W$- and $Z$-boson masses.  At tree level in the 
electroweak theory, we have
\begin{eqnarray}
    M_{W}^{2} & = & g^{2}v^{2}/2 = 
    \pi\alpha/G_{F}\sqrt{2}\sin^{2}\theta_{W} ,
    \label{eq:Wmass}  \\
    M_{Z}^{2} & = & M_{W}^{2}/\cos^{2}\theta_{W} ,
    \label{eq:Zmass}  
\end{eqnarray}
where the electroweak scale is $v = (G_{F}\sqrt{2})^{-\frac{1}{2}}
\approx 246\gev$.  But the electroweak scale is not the only scale.  
It seems certain that we must also consider the Planck scale, derived 
from the strength of Newton's constant,
\begin{equation}
    M_{\mathrm{Planck}} =  (\hbar c/G_{\mathrm{Newton}})^{\frac{1}{2}} \approx 1.22 \times 10^{19}\gev \; .
    \label{eq:planck}
\end{equation}
It is also probable that we must take account of the $SU(3)_{c}\otimes 
SU(2)_{L}\otimes U(1)_{Y}$ unification scale around 
$10^{15\mathrm{-}16}\gev$.  And, as we heard on the first day of the 
workshop from Antonio Masiero,\cite{am} there may well be a distinct 
flavor scale.  The existence of other scales is behind the famous 
problem of the Higgs scalar mass: how to keep the distant scales from 
mixing in the face of quantum corrections, or how to stabilize the 
mass of the Higgs boson on the electroweak scale.

It is because $G_{\mathrm{Newton}}$ is so small (or because 
$M_{\mathrm{Planck}}$ is so large) that we normally consider 
gravitation irrelevant for particle physics.  The 
graviton-quark-antiquark coupling is generically $\sim 
E/M_{\mathrm{Planck}}$, so it is easy to make a dimensional estimate 
of the branching fraction for a gravitationally mediated rare kaon 
decay: $B(K_{L} \to \pi^{0}G) \sim (M_{K}/M_{\mathrm{Planck}})^{2} \sim 
10^{-38}$, which is truly negligible!

We know from the electroweak theory alone that the 1-TeV scale is 
special.  Partial-wave unitarity applied to gauge-boson scattering 
tells us that unless the Higgs-boson mass respects
\begin{equation}
    M_{H}^{2} < \frac{8\pi\sqrt{2}}{3G_{F}} \approx 1\tev^{2} \; ,
    \label{eq:higgsu}
\end{equation}
new physics is to be found on the 1-TeV scale.  To stabilize the
Higgs-boson mass against uncontrolled quantum corrections, and to
resolve the mass-hierarchy problem, we consider electroweak physics
beyond the standard model.  The most promising approaches are to 
generalize $SU(3)_{c}\otimes SU(2)_{L}\otimes U(1)_{Y}$ to a theory 
with a composite Higgs boson in which the electroweak symmetry is
broken dynamically (technicolor and related theories) or to a
supersymmetric standard model.

Let us look a little further at the problem of fermion masses.  In the 
electroweak theory, the value of each quark or charged-lepton mass is 
set by a new, unknown, Yukawa coupling.  Taking the electron as a 
prototype, we define the left-handed doublet and right-handed singlet
\begin{equation}
    \mathsf{L} = \left( 
    \begin{array}{c}
        \nu_{e}  \\
        e
    \end{array}
    \right)_{L} \; , \qquad \mathsf{R} \equiv e_{R}.
    \label{eq:elec}
\end{equation}
Then the Yukawa term in the electroweak Lagrangian is
\begin{equation}
    \mathcal{L}_{\mathrm{Yukawa}}^{(e)} = - 
    \zeta_{e}[\bar{\mathsf{R}}(\varphi^{\dagger}\mathsf{L}) + 
    (\bar{\mathsf{L}}\varphi)\mathsf{R}] \; ,
    \label{eq:eYuk}
\end{equation}
where $\varphi$ is the Higgs field, so that the electron mass is
$m_{e} = \zeta_{e}v/\sqrt{2}$.  Inasmuch as we do not know how to
calculate the fermion Yukawa couplings $\zeta_{f}$, I believe that
\textit{we should consider the sources of all fermion masses as
physics beyond the standard model.}  

Note that the values of the Yukawa couplings are vastly different for
different fermions: for the top quark, $\zeta_{t} \approx 1$, for the
electron $\zeta_{e} \approx 3\times 10^{-6}$, and if the neutrinos
have Dirac masses, presumably $\zeta_{\nu} \approx
10^{-10}$.\footnote{I am quoting the values of the Yukawa couplings at
a low scale typical of the masses themselves.} What accounts for the
range and values of the Yukawa couplings?  Our best hope until now has
been the suggestion from unified theories that the pattern of fermion
masses simplifies on high scales.  The classic intriguing prediction
of the $SU(5)$ unified theory involves the masses of the $b$ quark and
the $\tau$ lepton, which are degenerate at the unification point for a
simple pattern of spontaneous symmetry breaking.  The different
running of the quark and lepton masses to low scales then leads to the
prediction $m_{b} \approx 3 m_{\tau}$, in suggestive agreement with
what we know from experiment.

The conventional approach to new physics has been to extend the 
standard model to understand why the electroweak scale (and the mass 
of the Higgs boson) is so much smaller than the Planck 
scale.\cite{am}  A novel approach that has been developed over the 
past two years is instead to \textit{change gravity} to understand 
why the Planck scale is so much greater than the electroweak scale.  
Now, experiment tells us that gravitation closely follows the 
Newtonian force law down to distances on the order of $1\mm$.  Below 
about a millimeter, the constraints on deviations from Newton's 
inverse-square law deteriorate rapidly, so nothing prevents us from 
considering changes to gravity even on a small but macroscopic scale.  
For its internal consistency, string theory requires an additional six 
or seven space dimensions, beyond the $3+1$ dimensions of everyday 
experience.  Until recently it has been presumed that the extra 
dimensions must be compactified on the Planck scale, with a
compactification radius $R_{\mathrm{unobserved}} \approx
1/M_{\mathrm{Planck}} \approx 1.6 \times 10^{-35}\m$.  The new wrinkle 
is to consider that the ($SU(3)_{c}\otimes SU(2)_{L}\otimes U(1)_{Y}$)
standard model gauge fields, plus needed extensions, reside on 
$3+1$-dimensional branes, not in the extra dimensions, but that 
gravity can propagate into the extra dimensions.

How does this hypothesis change the picture?  The dimensional 
analysis (Gauss's law, if you like) that relates Newton's constant to 
the Planck scale changes.  If gravity propagates in $n$ extra 
dimensions with radius $R$, then
\begin{equation}
    G_{\mathrm{Newton}} \sim M_{\mathrm{Planck}}^{-2} \sim M^{\star\,-n-2}R^{-n}\; ,
    \label{eq:gauss}
\end{equation}
where $M^{\star}$ is gravity's true scale.  Notice that if we boldly 
take $M^{\star}$ to be as small as $1\tev$, then the radius of the extra 
dimensions is required to be smaller than about $1\mm$, for $n \ge 
2$.  What we know as the Planck scale is then a mirage that results 
from a false extrapolation: treating gravity as four-dimensional down 
to arbitrarily small distances, when in fact---or at least in this 
particular fiction---gravity propagates in $3+n$ spatial dimensions.  
The Planck mass is an artifact, given by $M_{\mathrm{Planck}} = 
M^{\star}(M^{\star}R)^{n/2}$.

Although the idea that extra dimensions are just around the 
corner---either on the submillimeter scale or on the TeV scale---is 
preposterous, it is not ruled out by observations.  For that reason 
alone, we should entertain ourselves by entertaining the 
consequences.  Many authors have considered the gravitational 
excitation of a tower of Ka\l uza--Klein modes in the extra 
dimensions, which would give rise to a missing (transverse) energy 
signature in collider experiments.\cite{smaria}  We call these excitations 
\textit{provatons,} after the Greek word for a sheep in a 
flock.\footnote{I thank Maria Spiropulu for instructing me in the 
difference between $\pi\rho\acute{o}\beta\alpha\tau o\nu$ and 
$\alpha\rho\nu\grave{\iota}$.}

``Large'' extra dimensions present us with new ways to think about the 
exponential range of Yukawa couplings.  Arkani-Hamed, Schmaltz, and 
collaborators\cite{martin} have speculated that if the standard-model 
brane has a small thickness, the wave packets representing different 
fermion species might have different locations within the extra 
dimension.  On this picture, the Yukawa couplings measure the overlap 
in the extra dimensions of the left-handed and right-handed wave 
packets and the Higgs field, presumed pervasive.  Exponentially large 
differences might then arise from small offsets in the new 
coordinate(s).  True or not, it is a completely different way of 
looking at an important problem.

\section{CP Violation in the Standard Model}
In the standard electroweak theory, the charged-current interactions 
may be represented as
\begin{equation}
    \left(
    \begin{array}{ccc}
        \bar{u} & \bar{c} & \bar{t}
    \end{array}
    \right)_{L} \mathsf{V} \left( 
    \begin{array}{c}
        d  \\
        s  \\
        b
    \end{array}
    \right)_{L} \; ,
    \label{eq:smcc}
\end{equation}
where the (Cabibbo--Kobayashi--Maskawa\cite{nicola,km}) quark-mixing matrix is
\begin{equation}
    \mathsf{V} = \left( 
    \begin{array}{ccc}
        V_{ud} & V_{us} & V_{ub}  \\
        V_{cd} & V_{cs} & V_{cb}  \\
        V_{td} & V_{ts} & V_{tb}
    \end{array}
    \right)\; .
    \label{eq:ckm}
\end{equation}
Within the framework of the standard model, the elements of 
$\mathsf{V}$ originate in the Yukawa couplings of the quarks to the 
Higgs field; accordingly, they offer a link to physics beyond the 
standard model.  For three generations of quarks, the mixing matrix 
depends on three real angles and one phase; the phase is the source 
of \CP\ violation.\cite{smcp,ali,guidetto}

The quark-mixing matrix is unitary: $\mathsf{V}\mathsf{V}^{\dagger} =
\mathsf{I}$.  Consequently the product of any row or column of the
matrix with the complex conjugate of another must vanish.  For a $3\times 3$
matrix, each such equation may be depicted as a closed triangle in the
complex plane.  There are six distinct unitarity
triangles.\cite{bigisan} In a convenient
parametrization,\cite{lincoln} we can express the quark-mixing matrix
in terms of three real quantities and one imaginary quantity as
\begin{equation}
   \mathsf{V} = \left( 
   \begin{array}{ccc}
       1 - \cfrac{1}{2}\lambda^{2} & \lambda & A\lambda^{3}(\rho - 
       i\eta)  \\
       -\lambda & 1 - \cfrac{1}{2}\lambda^{2} & A\lambda^{2}  \\
       A\lambda^{3}(1 - \rho - i\eta) & -A\lambda^{2} & 1
   \end{array}
   \right) + \mathcal{O}(\lambda^{4}) \; ,
    \label{eq:wolf}
\end{equation}
The phases lie in the extreme off-diagonal elements; they are 
designated by
\begin{equation}
    \left( 
    \begin{array}{ccc}
        1 & 1 & e^{-i\gamma}  \\
        1 & 1 & 1  \\
        e^{-i\beta} & 1 & 1
    \end{array}
    \right)\; .
    \label{eq:ckmph}
\end{equation}
A third angle is defined by the triangle constraint $\alpha \equiv 
\pi - \beta - \gamma$.

\begin{figure}[tb]
\centerline{\BoxedEPSF{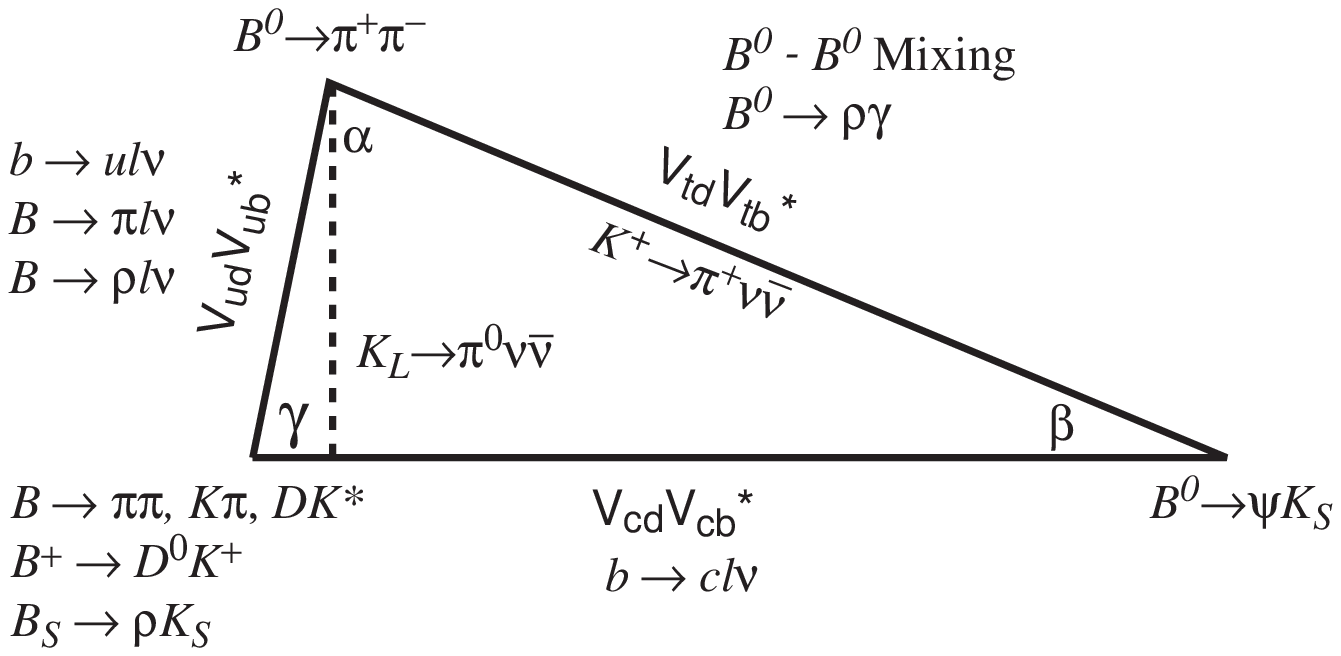 scaled 475} 
\BoxedEPSF{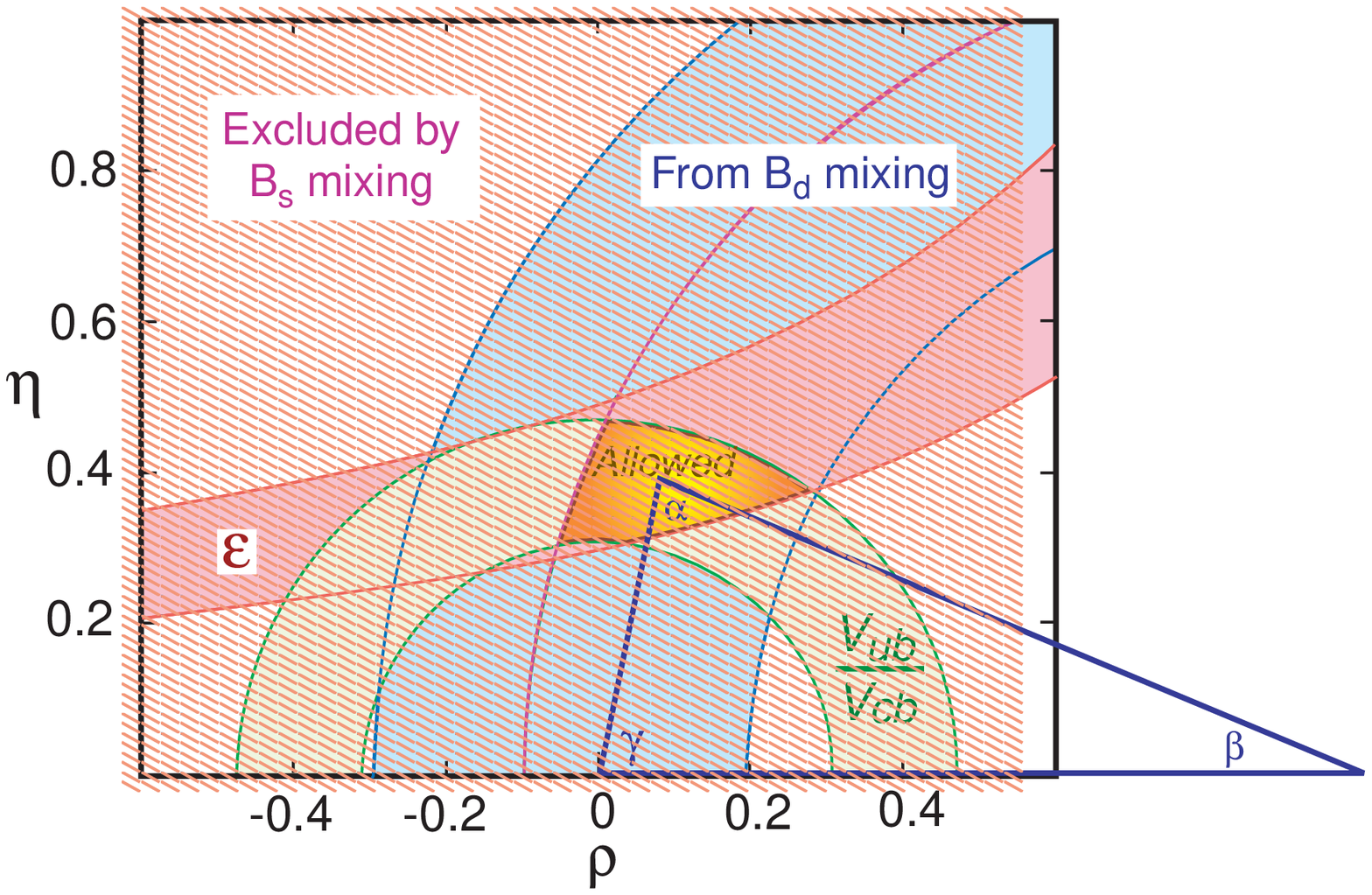 scaled 475}}
\vspace*{6pt}
\caption{\textit{The $bd$ unitary triangle.  On the left, the angles 
$\alpha$, $\beta$, and $\gamma$ are defined and some of the 
decay modes that allow measurements of the sides, angles, and altitude 
are indicated.  On the right, the normalized unitary triangle is 
depicted in the $\rho$-$\eta$ plane, and the current experimental 
constraints are indicated. }
    \label{fig:triangles}}
\end{figure}

One way to characterize the task of testing the hypothesis that all 
\CP\ violation (among the known particles) arises from the phase in 
the quark-mixing matrix is to try to overconstrain the unitarity 
triangle(s), by measuring the sides and angles in many different 
ways.  The $bd$ triangle is shown twice in Figure 
\ref{fig:triangles}.  On the left, I show the triangle schematically, 
and indicate the kinds of measurements that give us information about 
the various sides and angles.  
On the right, I show a summary of what we currently know about the
shape of the $bd$ triangle.\cite{sheldon} The sources of our knowledge
were reviewed in Frascati by Ahmed Ali,\cite{ali} Guido
Martinelli,\cite{guidetto}, and Gerhard Buchalla\cite{gerhard}.  Our
most important objectives for the near future are the measurements of
the quantity $\sin2\beta$ in $B^{0} \to J/\psi\;K_{S}$ decays and the
rate of $B_{s}\hbox{--}\bar{B}_{s}$ mixing, and pushing on to measure
the rates for the \textit{rarissime} decays $K^{+} \to
\pi^{+}\nu\bar{\nu}$ and $K_{L} \to \pi^{0}\nu\bar{\nu}$.

Before leaving these generalities, I want to offer a few comments 
about \CP\ violation and the matter asymmetry of the Universe.  We 
learned long ago from Andrei Sakharov\cite{as} that three criteria 
must be fulfilled to generate a matter asymmetry in a universe that 
begins from neutral initial conditions.  We require
\begin{enumerate}
    \item  Microscopic \CP\ violation, such as Christenson, \etal\ 
    first observed in the decay $K_{L} \to \pi\pi$.\cite{ccft}

    \item  Baryon- (lepton-) number violating processes, of the kind 
    implied by unified theories of the strong, weak, and 
    electromagnetic interactions.

    \item  A universe that evolves out of thermal equilibrium during 
    baryogenesis, a condition natural in the hot big-bang scenario.
\end{enumerate}
According to our current understanding of baryogenesis,\cite{at} the
observed baryon excess in the Universe, characterized by the
baryon-to-photon ratio $n_{B} / n_{\gamma} \approx 4 \times 10^{-10}$,
cannot be reproduced within the standard model by the \CP\ violation
that arises from the quark-mixing phase.  It is just barely possible
to generate the observed baryon density at the electroweak scale
within the minimal supersymmetric extension of the standard model, but
this requires that the lightest Higgs boson lie within reach of LEP
experiments, $M_{h} \ltap 105\gevcc$, \textit{and} that the lighter
top squark be less massive than the top quark, $m_{\tilde{t}_{1}}
\ltap m_{t}$.  These two requirements do not yet conflict
with experiment, but it is likely that we need new sources of \CP\
violation to account for the matter excess.  It is important that we
speak precisely when we tell the world what we expect to learn from 
the detailed study of \CP\ violation in the $B$ mesons: it is 
unlikely that we shall unlock the secrets of primordial baryogenesis, 
although the increased understanding we gain into \CP\ violation in 
the quark sector may give us new insights into what it would take to 
explain the matter excess.  We should not promise what we do not 
expect to deliver!\newpage

\section{Workshop Headlines}

\subsection{$\epsp \neq 0$!}
The most significant result of this year for the study of \CP\ 
violation is the conclusion from the KTeV experiment at Fermilab and 
the NA48 experiment at CERN that \CP\ invariance is violated directly 
in the decay $K_{L} \to \pi\pi$.  Let us recall for a moment the 
phenomenological setting of these new experimental results.

In a \textsf{CPT}-invariant theory, we can write the neutral kaons of 
definite mass and lifetime as
\begin{equation}
    \ket{K_{S}} = p\ket{K^{0}} + q\ket{\bar{K}^{0}}\; , \quad
    \ket{K_{L}} = p\ket{K^{0}} - q\ket{\bar{K}^{0}}\; .
    \label{eq:ksldefs}
\end{equation}
If \CP\ invariance holds, then $p = q = 1/\sqrt{2}$, and 
$(\ket{K_{S}},\ket{K_{L}})$ is a \CP\ eigenstate with \CP\ 
$=(+1,-1)$.  A small \CP\ impurity can be represented by the 
parameter $\varepsilon$, through the connection
\begin{equation}
    \frac{p}{q} = \frac{(1 + \varepsilon)}{(1 - \varepsilon)}\; ;
    \label{eq:pbyq}
\end{equation}
the observed $K_{L} \to \pi\pi$ decay rate fixes $|\varepsilon| 
\approx 2.28 \times 10^{-3}$.  A direct \CP\ violation in the decay 
amplitude leads to unequal rates for the charged and neutral decay 
modes that is expressed through the parameter $\epsp$, as\cite{wuyang}
\begin{eqnarray}
    \eta_{+-} = & 
    \displaystyle{\frac{A(K_{L}\to\pi^{+}\pi^{-})}{A(K_{S}\to\pi^{+}\pi^{-})}}
    & = \varepsilon + \epsp \; ,
    \label{eq:etapm}  \\
    \eta_{00} =  & 
    \displaystyle{\frac{A(K_{L}\to\pi^{0}\pi^{0})}{A(K_{S}\to\pi^{0}\pi^{0})}} 
    & = \varepsilon - 2\epsp \; .
    \nonumber
\end{eqnarray}
The parameter $\varepsilon$, which measures the \CP-impurity in the 
$\ket{K_{L}}$ and $\ket{K_{S}}$ states, arises in the standard model 
from box diagrams, whereas the parameter $\epsp \ll \varepsilon$, 
which measures direct \CP\ violation, arises from penguin diagrams.  
In the standard model, $\real{\epsp/\varepsilon} \approx 10^{-3}$; in 
the superweak phenomenology,\cite{sweak} $\real{\epsp/\varepsilon} = 0$.

A comparison of the charged and neutral decay rates yields a measure 
of $\epsp/\varepsilon$ as
\begin{equation}
    |\eta_{+-}|^{2}/|\eta_{00}|^{2} \approx 1 + 6 \real{\epsp/\varepsilon}\;.
    \label{eq:epmzz}
\end{equation}
The decisive new results on $\real{\epsp/\varepsilon}$ reported within 
the past year are
\begin{center}
    \begin{tabular}{cl}
        KTeV\cite{cheu,ktevpub}: & $(28.0 \pm 4.1) \times 10^{-4}$, and \\
        NA48\cite{duclos,na48pub}: & $(18.5 \pm 7.3) \times 10^{-4}$,  \\
    \end{tabular}
\end{center}
which are in good agreement with each other, and with the earlier 
CERN result from NA31,\cite{na31} $\real{\epsp/\varepsilon} = (23 \pm 
6.5) \times 10^{-4}$.  The KTeV measurement in particular is not in 
close agreement with the earlier measurement from Fermilab experiment 
E731,\cite{e731} $\real{\epsp/\varepsilon} = (7.4 \pm 5.9) \times 
10^{-4}$.  Taken together, all the results lead to a world average
\begin{equation}
    \real{\epsp/\varepsilon} = (21.3 \pm 2.8) \times 10^{-4}\; ,
    \label{eq:epspavg}
\end{equation}
which is convincingly different from zero.  We can therefore draw the 
important conclusion that \textit{the superweak picture does not explain all} 
\CP\ \textit{violation.}

In my view, these two experiments are among the most beautiful we have 
in particle physics today.  As we heard in the talks by Elliott 
Cheu\cite{cheu} and Jean Duclos,\cite{duclos}, both aim in the future 
for a precision of $\pm 1 \times 10^{-4}$.  That same level of 
precision is the target for \textsc{kloe,} in its special setting at 
the $\phi(1019)$ resonance, and we saw in the presentation by 
Matteo Palutan\cite{kloe} a clean \CP-violating signal from the \textsc{kloe} 
shakedown cruise.  We look forward with anticipation to the first 
high-luminosity running of \dafne.  The Budker Institute in 
Novosibirsk has its own designs on a $\phi$ factory with a 
luminosity of $2.5 \times 10^{33}\lum$, and for VEPP-2000, which is 
intended to investigate particle production up to $2\gev$ in regions 
not exhaustively studied.\cite{novo}

What more can we make of the value of $\epsp/\varepsilon$ now in hand, 
and of the exquisitely precise results to come?  It is fair to say 
that the new KTeV result seemed at first sight shockingly large; the 
expectation in nearly everyone's head was for a number no longer than 
$10$ (in $10^{-4}$ units).  Perhaps while awaiting the new results we 
had all become mesmerized by the dependence of $\epsp/\varepsilon$ 
upon $m_{t}$ and the threat that a heavy top quark might mean little 
discrimination between the three-generation electroweak theory and the 
superweak hypothesis.  Surely we should have taken more note (before 
the event) of the $m_{s}^{-2}$-sensitivity of (the usual 
parametrization of) $\epsp/\varepsilon$ and the trend in modern 
lattice determinations toward smaller strange-quark 
masses.\cite{sinead} Perhaps hearing similar predictions year after 
year for $\epsp/\varepsilon$ we lost a sense of how difficult is the 
step from quarks to hadrons and accorded too much respect to the 
central values.  In any event, the KTeV value provoked many 
reexaminations of how well we can predict $\epsp/\varepsilon$ in the 
standard model,\cite{tmu,andrzej2,ulin,fabbr,di12,garval,manny} as 
well as searches\cite{hitoshi} for evidence of new physics in the 
value of $\epsp/\varepsilon$.  A new lattice study by the 
Brookhaven--Columbia--RIKEN group\cite{tblum} that yields a sign 
opposite to other calculations (and the data!) has not settled the 
matter.  Guido Martinelli's conclusion is apt: Calculating 
$\epsp/\varepsilon$ ``requires control of strong-interaction dynamics 
at a level of accuracy still not reached by the 
theory.''\cite{guidetto}

\subsection{CDF's Measurement of $\sin2\beta$}
We do not yet have a definitive observation of \CP\ violation in $B$ 
decays, but the CDF Collaboration has made impressive progress toward 
that goal.\cite{cb}  In their 110-pb$^{-1}$ Run 1 at the Tevatron 
Collider, they have reconstructed a sample of $395 \pm 31$ $B^{0} 
\to \psi K_{S}$ events.  To construct the \CP-violating asymmetry for 
the decay rates of $B^{0}$ and $\bar{B}^{0}$ into this common decay 
mode, it is necessary to tag the flavor of the neutral $B$ meson at 
the moment of its birth.  CDF has employed both an opposite-side tag, 
using the charge of a soft lepton or a jet-charge algorithm to 
identify the flavor of the $b$ or $\bar{b}$ produced in association 
with the state that decays into $\psi K_{S}$.  They have also used 
the charge of a nearby hadron to tag the decaying object itself.  
Combining all their tagging methods, they arrive at a measurement
\begin{equation}
    \sin2\beta = 0.79 ^{+0.41}_{-0.44} \; ,
    \label{eq:cdfsin2b}
\end{equation}
which is non-negative at 93\% CL.  This does not constitute a proof 
that \CP\ is violated in $B$ decays, nor a precision measurement of 
$\sin2\beta$, but it is a proof of the method and the harbinger of 
things to come.  In Run 2 of the Tevatron Collider, a 2-fb$^{-1}$ 
exposure to begin in March 2001, they expect to reconstruct about 
$10^{4}$ $B^{0} \to \psi K_{S}$ events, and to measure $\sin2\beta$ 
with an uncertainity $\delta\sin2\beta \ltap 0.08$.  Other 
improvements---beyond their longer silicon vertex detector and the 
increased luminosity---may make possible a still better 
determination.  The D\O\ experiment will have a silicon vertex 
detector for the first time in Run 2.  They project an uncertainty 
of $\delta\sin2\beta \approx 0.15$.

The next run of CDF should also yield our first measurement of the 
frequency of $B_{s}\hbox{ - }\bar{B}_{s}$ mixing.  In experiments 
carried out so far (ALEPH, CDF, DELPHI, OPAL, and SLD), the 
oscillations are too rapid to be observed.  The combined results give 
the mass difference between the two $B_{s}\hbox{ - }\bar{B}_{s}$ 
mixtures as $\Delta m_{s} > 14.3\ps^{-1}$ at 95\% CL, or $x_{s} 
\equiv \tau_{s}\Delta m_{s} > 22.$  We currently project the 
standard-model value for the oscillation frequency to be no more than 
about $20\ps^{-1}$.  In Run 2, CDF expects to 
reconstruct about $20\,000$ examples of $B_{s} \to D_{s}(\pi, 
3\pi)$.  They should be sensitive to $x_{s} \ltap 63$, or to $\Delta 
m_{s} \ltap 41\ps^{-1}$.

In Run 2 of the Tevatron and beyond, CDF and D\O\ anticipate a very 
rich harvest of $B$ physics,\footnote{Consult the web pages of the 
ongoing Run 2 $b$ Workshop at Fermilab, for which links may be found 
at \textsf{http://www-theory.fnal.gov.}} including detailed studies of 
$B_{s}$, $B_{c}$, and $b$-baryons.  To quote one simple figure of 
merit, in an extended run of $(20, 30)\fb^{-1}$, CDF expects 
$\delta\sin2\beta = (0.03, 0.02)$.

Prospects are very bright for the hadron collider $B$ experiments that 
will succeed CDF and D\O, as we heard in detail from Ramon 
Miquel.\cite{ramon}  Again to give a simple measure of the power of 
these experiments, I show in Table \ref{tab:hcs2b} the quality of 
the $\sin2\beta$ measurements anticipated in one year of running.
\begin{table}
\centering
\caption{\textit{Precision expected in one year's running}}
\vskip 0.1 in
\begin{tabular}{|c|c|} \hline
Experiment & $\delta\sin2\beta$ \\
\hline
\hline
ATLAS & 0.021 \\
CMS & 0.025 \\
LHC$b$ & 0.015 \\
BTeV & 0.021 \\
\hline
\end{tabular}
\label{tab:hcs2b}
\end{table}
I refer to Miquel's talk for the capabilities of these experiments to 
measure $\alpha$, $\gamma$, and other quantities.  As the new hadron 
collider experiments prepare, the extremely challenging (fixed-target) 
\textsc{hera-}$B$ Experiment gives a preview of LHC-like running 
conditions.\cite{ida}

\subsection{CLEO Observes $B^{0} \to \pi^{+}\pi^{-}$}
After several years of publishing ever smaller upper bounds to the 
branching ratio for the rare decay $B^{0} \to \pi^{+}\pi^{-}$, the 
CLEO Collaboration now reports the definitive observation of this 
potentially important mode.\cite{davec}  The branching fraction 
\begin{equation}
    B(B^{0} \to \pi^{+}\pi^{-}) = (4.3 \pm 1.5 \pm 0.5) \times 10^{-6}
    \label{eq:Bpipi}
\end{equation}
is smaller than hoped, which will complicate both the observation and 
the interpretation of \CP\ violation in the $\pi\pi$ channel.  CLEO 
has also reported the first observation, at about the $10^{-5}$ level, 
of the rare decays $B^{+} \to K^{+} \pi^{0}$ and $B^{0} \to K^{0} 
\pi^{0}$.

CLEO has also made a new measurement of $B^{0}\hbox{--}\bar{B}^{0}$ 
mixing, determining
\begin{equation}
    \Delta m_{d} = (0.519 \pm 0.025 \pm 0.032)\ps^{-1}\; ,
    \label{eq:delmd}
\end{equation}
in excellent agreement with the world average.  Finally, to search 
for unexpected sources of \CP\ violation that might be needed to 
account for the baryon number of the Universe, CLEO has looked for 
\CP-violating asymmetries in eight rare decays.  The experimental 
uncertainties are larger than the asymmetries expected if the phase of 
the quark-mixing matrix is the source of the \CP\ violation, and no 
significant asymmetries are observed.

\subsection{\textsc{BaBar} and \textsc{belle} Are Coming}
I should perhaps entitle this paragraph, ``\textsc{BaBar} and
\textsc{belle} Are Here!'', for KEK-B and \textsc{belle} and PeP-II
and \textsc{BaBar} have completed their construction and commissioning
in very rapid and impressive fashion.  The performance of these two
asymmetric $B$ factories and their detectors, summarized here at
\dafne99 by Ida Peruzzi,\cite{ida} has been inspiring, and gladdens
the heart of every particle physicist.  Both experiments are already
producing quasi-physics plots, and we can reasonably hope for new
physics by the 2000 summer conferences.

The PeP-II collider\cite{pep2} has already reached half its design 
luminosity, with
\begin{equation}
    \mathcal{L}_{\mathrm{peak}} = 1.35 \times 10^{33}\lum\; .
    \label{eq:pep2lum}
\end{equation}
The \textsc{BaBar} detector has already logged more than $1\fb^{-1}$, 
and the experimenters will try to accumulate $10\fb^{-1}$ by next 
summer.  The PeP-II designers are planning for a threefold luminosity 
increase in 2002, and are beginning to work toward a tenfold increase 
in 2005.  The KEK-B machine has also made a good 
beginning,\cite{kekb} and luminosity is improving there.  The plan at 
KEK is now to run \textsc{belle} continuously for about ten months.

\subsection{$K \to \pi\nu\bar{\nu}$}
One of the outstanding recent results was the observation at 
Brookhaven of the first candidate---a superb candidate, at that---for 
the rare decay $K^{+} \to \pi^{+}\nu\bar{\nu}$ in Experiment 
787.\cite{e787pub} As I wrote at the time, ``What is most impressive 
to me is not the one beautiful candidate event, but the extremely low 
level of background: the event occurs on an empty field.''\cite{hq98} 
Given the sensitivity at the time, that one event led to a branching 
fraction $\Gamma(K^{+} \to \pi^{+}\nu\bar{\nu})/\Gamma(K^{+} 
\to\hbox{all}) = (4.2^{+9.7}_{-3.5})\times 10^{-10}$, consistent with 
the standard-model expectation\cite{gerhard} of $(0.8 \pm 0.3) \times 
10^{-10}$, but tantalizingly large.  As we heard in Frascati from 
Takahiro Sato,\cite{sato} a trebled data set has led to no new candidate 
events, still with an admirable absence of background near the signal 
region.  The new (and preliminary) branching fraction is
\begin{equation}
    B(K^{+} \to \pi^{+}\nu\bar{\nu}) = 1.5^{+3.5}_{-1.3} \times 10^{-10} \;,
    \label{eq:e787br}
\end{equation}
which could hardly agree better with the standard model.  

Because of the clean theoretical interpretation of the $K^{+} \to
\pi^{+}\nu\bar{\nu}$ mode and its potential sensitivity to new
physics, it is clearly of interest to subject the standard model to a
more incisive test.  There are plans at Brookhaven for an evolution of 
E787 to a project known as E949, which promises a single-event 
sensitivity $B(K^{+} \to \pi^{+}\nu\bar{\nu}) = 1.4 \times 10^{-11}$, 
with an expected background of 0.7 event.  By adding the kinematic 
region below the $K^{+} \to \pi^{+}\pi^{0}$ line, the E949 
experimenters project an ultimate sensitivity $B(K^{+} \to 
\pi^{+}\nu\bar{\nu}) = 0.8 \times 10^{-11}$.\cite{e949}  An experiment currently 
under study at Fermilab as an R\&D project, CKM, aims for a 10\% 
measurement of $B(K^{+} \to \pi^{+}\nu\bar{\nu})$.\cite{ronray,ckmprop}

The neutral analogue decay, $K_{L} \to \pi^{0}\nu\bar{\nu}$, for which
the standard-model branching fraction is $(2.8 \pm 1.1) \times
10^{-11}$,\cite{gerhard} directly measures the altitude $\eta$ of the $bd$
unitarity triangle.  Three experiments plan to take on the very 
considerable challenge of detecting $K_{L} \to \pi^{0}\nu\bar{\nu}$.  
At KEK, Experiment E391 will begin in 2001 at the existing 12-GeV proton 
synchrotron, where a sensitivity at the $10^{-10}$ level is 
anticipated.  If all goes well, it would evolve into a more sensitive 
experiment at the Japanese Hadron Facility.  Brookhaven has approved, 
but not yet secured funding for, E926, or \textsc{kopio}, which would 
record about 50 events at the standard-model rate.\cite{bryman,e926}  At 
Fermilab, the next phase of the KTeV program would be the \textsc{kami} (Kaons 
At the Main Injector) experiment, currently an R\&D project.\cite{kami}  Its goal 
is a 10\% measurement of $\eta$ and sensitivity to other rare decays 
at the $10^{-13}$ level.\cite{ronray}

Both $K^{+}$ and $K_{L}$ experiments are fantastically challenging,
but they promise indispensable tets of the three-generation
electroweak theory.\cite{am,gerhard,nirworah,ajbetal,gino} To balance
the world's research portfolio---so heavily invested in the $B$
sector---we need to pursue the $K \to \pi \nu\bar{\nu}$ channels
vigorously.

\subsection{Persistent Electric Dipole Moments}
The electric dipole moment of a particle or structure is defined in 
terms of its charge distribution $\rho(\vec{x})$ as
\begin{equation}
    \vec{D} = \int d^{3}\vec{x}\;\vec{x}\rho(\vec{x})\; ,
    \label{eq:edmdef}
\end{equation}
which must be directed along (or opposite to) the spin direction 
$\vec{s}$, because the spin is the only directional reference the 
particle carries.  Under the action of a parity inversion, \textsf{P}, 
the electric dipole moment is reversed ($\vec{D} \to - \vec{D}$), 
while the spin (pseudo)vector is unchanged ($\vec{s} \to \vec{s}$).  
Under time reversal, \textsf{T}, the electric dipole moment is 
unchanged ($\vec{D} \to \vec{D}$), while the spin direction is 
reversed ($\vec{s} \to - \vec{s}$).  Accordingly, the persistent 
electric dipole moment must vanish unless both \textsf{P} and 
\textsf{T} are violated.  A nonvanishing electric dipole moment 
therefore implies \textsf{T} violation and, in a 
\textsf{CPT}-invariant world, \CP\ violation.

Since the discovery of \CP\ violation in $K_{L}$ decays in 1964, the 
neutron electric dipole moment has been the target of many 
experiments, first with neutron beams and later with ultracold 
neutrons.\cite{pdg}  As Mike Pendlebury reminded us,\cite{edms} the 
experimental upper bounds have dropped steadily, sweeping away a 
number of theoretical speculations.  Whether or not persistent 
electric dipole moments exist, we can be grateful for the persistent 
experimenters who are hunting them down.  An improved limit reported 
this year from the Institut Laue-Langevin, Grenoble,\cite{dn99} 
\begin{equation}
    |d_{n}| < 6.3 \times 10^{-26}\;e\cm\hbox{ at 90\% CL},
    \label{eq:dn99}
\end{equation}
is still about six orders of magnitude greater than the standard-model 
expectation, but menaces multi--Higgs-boson models for \CP\ violation, 
and approaches the predictions of supersymmetric models.  The ILL 
group anticipates an improvement by a factor of 3 over about three 
years; an improvement by two orders of magnitude may be possible over 
a decade.

The past decade has brought remarkable progress in the search for an 
electric dipole moment of the electron.  The most sensitive measurement 
is obtained using the amplification of $d_{e}$ in atomic 
thallium,\cite{de94}
\begin{equation}
    d_{e}= (1.8 \pm 1.2 \pm 1.0) \times 10^{-27}\;e\cm\;,
    \label{eq:de94}
\end{equation}
which leads to an upper limit of
\begin{equation}
    |d_{e}| \ltap 4 \times 10^{-27}\;e\cm\; .
    \label{eq:de94lim}
\end{equation}
That is impressive in absolute terms, but---because of the electron's 
small mass---at least ten orders of magnitude greater than the range 
predicted in the standard model.  The good news, of course, is that a 
lot of terrain is open for an important discovery.  A new technique 
using the YbF molecule to amplify the effect of $d_{e}$ may lead to a 
tenfold increase in sensitivity over three years, and perhaps another 
order of magnitude beyond that is in prospect.

The current published limit on the muon's electric dipole moment dates 
to the classic $(g-2)_{\mu}$ experiment at CERN.\cite{cerngm2}  That 
number,
\begin{equation}
    |d_{\mu}| \ltap 7 \times 10^{-19}\;e\cm\; ,
    \label{eq:dmu}
\end{equation}
should be improved by about a factor of twenty in the $(g-2)_{\mu}$ 
experiment under way at Brookhaven.\cite{gm2}  A proposed dedicated 
experiment at Brookhaven\cite{muedm} could improve the sensitivity to
\begin{equation}
    |d_{\mu}| \approx 10^{-24}\;e\cm.
    \label{eq:bnldmu}
\end{equation}
That would be as sensitive, in relative terms, as the current limit 
on the electron's electric dipole moment, since we expect $|d_{\mu}|
\approx (m_{\mu}/m_{e})|d_{e}|$.

\subsection{The Search for Lepton Flavor Violation}
The observation of lepton flavor violation would be a clear sign of 
physics beyond the standard electroweak theory.  High-sensitivity 
experiments provide access to high mass scales for the particles that 
might mediate lepton flavor violation.  Although these two facts are 
reason enough to pursue the search for lepton flavor violation, the 
experimental evidence for neutrino oscillations has renewed interest in 
LFV searches, although it must be said that the connection is highly 
model dependent.\cite{kunokada}

For more than four decades, the upper bounds on a variety of 
lepton-flavor-violating decay modes have decreased by a factor of ten 
every seven years.\cite{yoshi}  In the realm of kaon physics, the 
decay $K_{L} \to \mu^{\pm}e^{\mp}$ is sensitive to axial and 
pseudoscalar LFV operators, while the decays $K^{+} \to 
\pi^{+}\mu^{\pm}e^{\mp}$ and $K_{L} \to \pi^{0}\mu^{\pm}e^{\mp}$ 
probe vector and scalar LFV operators.  Brookhaven experiment E871's
current limit on the purely leptonic decay mode,\cite{e871}
\begin{equation}
    B(K_{L} \to \mu^{\pm}e^{\mp}) < 4.7 \times 10^{-12}\hbox{ (90\% CL)},
    \label{eq:kmue}
\end{equation}
is the smallest limit on the branching fraction of a hadron.  
Experiment E865 at Brookhaven has improved its limit on the $K^{+} 
\to \pi^{+}\mu^{\pm}e^{\mp}$ branching fraction to
\begin{equation}
    B(K^{+}\to \pi^{+}\mu^{\pm}e^{\mp}) < 2.9 \times 10^{-11}\hbox{ (90\% CL)}.
    \label{eq:Kppimue}
\end{equation}
With five times more data accumulated in 1998, they project a 90\%-CL 
sensitivity of $6\times 10^{-12}$.  The corresponding neutral-kaon 
branching fraction, reported by Fermilab experiment E799 
(KTeV),\cite{e799} is
\begin{equation}
    B(K_{L} \to \pi^{0}\mu^{\pm}e^{\mp}) < 3.1 \times 10^{-9}\; .
    \label{eq:K0pimue}
\end{equation}

Many experiments have now used natural sources of neutrinos, neutrino 
radiation from fission reactors, and neutrino beams generated in 
particle accelerators to look for evidence of neutrino oscillation.  
The positive indications for neutrino oscillations fall into three 
classes:\cite{janetc}
\begin{enumerate}
	\item  Five solar-neutrino experiments report deficits with respect 
	to the predictions of the standard solar model: Kamiokande and 
	Super-Kamiokande using water-Cherenkov techniques, SAGE and GALLEX 
	using chemical recovery of germanium produced in neutrino 
	interactions with gallium, and Homestake using radiochemical 
	separation of argon produced in neutrino interactions with 
	chlorine.  These results suggest the oscillation $\nu_{e} 
	\rightarrow \nu_{x}$, with $|\Delta m^{2}|_{\mathrm{solar}} \approx 
	10^{-5}\ev^{2}$ and $\sin^{2}2\theta_{\mathrm{solar}}\approx 1\hbox{ 
	or a few}\times 10^{-3}$, or $|\Delta m^{2}|_{\mathrm{solar}} \approx 
	10^{-10}\ev^{2}$ and $\sin^{2}2\theta_{\mathrm{solar}}\approx 1$.

	\item Five atmospheric-neutrino experiments report anomalies in the
	arrival of muon neutrinos: Kamiokande, IMB, and Super-Kamiokande using
	water-Cherenkov techniques, and Soudan II and MACRO using sampling
	calorimetry.  The most striking result is the zenith-angle dependence
	of the $\nu_{\mu}$ rate reported last year by Super-K
	\cite{SKatm,SKLyon}.  These results suggest the oscillation $\nu_{\mu}
	\rightarrow \nu_{\tau}\hbox{ or }\nu_{s}$, with
	$\sin^{2}2\theta_{\mathrm{atm}} \approx 1$ and $|\Delta
	m^{2}|_{\mathrm{atm}} = 10^{-3}\hbox{ to }10^{-4}\ev^{2}$.

	\item The LSND experiment \cite{LSND} reports the observation of
	$\bar{\nu}_{e}$-like events is what should be an essentially pure
	$\bar{\nu}_{\mu}$ beam produced at the Los Alamos Meson Physics
	Facility, suggesting the oscillation $\bar{\nu}_{\mu} \rightarrow
	\bar{\nu}_{e}$.  This result has not yet been reproduced by any other
	experiment.  The favored region lies along a band from
	$(\sin^{2}2\theta_{\mathrm{LSND}} = 10^{-3},|\Delta
	m^{2}|_{\mathrm{LSND}} \approx 1\ev^{2})$ to 
	$(\sin^{2}2\theta_{\mathrm{LSND}} = 1,|\Delta
	m^{2}|_{\mathrm{LSND}} \approx 7 \times 10^{-2}\ev^{2})$. 
\end{enumerate}
A host of experiments have failed to turn up evidence for neutrino 
oscillations in the regimes of their sensitivity.  These results limit 
neutrino mass-squared differences and mixing angles.  In more than a 
few cases, positive and negative claims are in conflict, or at least 
face off against each other.  Over the next five years, many 
experiments will seek to verify, further quantify, and extend these 
claims.

Groups at Berkeley, Brookhaven, CERN, Fermilab, and KEK are
investigating the feasibility of using muon decay rings as intense
neutrino sources.  Studies directed toward the eventual construction
of $\mu^{+}\mu^{-}$ colliders suggest that it may be possible to
accumulate approximately a millimole of muons per year.  What is a bug
for the muon colliders---the decay $\mu^{-} \to e^{-}
\nu_{\mu}\bar{\nu}_{\mu}$---becomes a feature for a neutrino
factory.\footnote{For an overview, see the web site for NuFact '99, 
the ICFA/ECFA Workshop, ``Neutrino Factories based on Muon Storage 
Rings,'' held last summer in Lyon, at 
\textsf{http://lyoinfo.in2p3.fr/nufact99/.}} With a stored muon beam 
with an energy of tens of GeV, it may be practical to illuminate a 
distant detector virtually anywhere on Earth and to have an event rate 
useful for oscillation studies.\cite{sgeer}  Under rather special circumstances, 
it may be possible to observe \CP\ violation in neutrino 
oscillations.\cite{nucp} The prospect of intense high-energy beams of 
electron neutrinos and antineutrinos is very intriguing.  For now, the 
principal questions are whether a neutrino factory is feasible and 
whether it is the right instrument to address the next-generation 
questions in neutrino mass and mixing.  I expect useful first-order 
answers to these questions within a year.

\subsection{\textsf{T} Violation in $K$ 
Decays}
About a year ago, the CPLEAR Collaboration at CERN reported on the
first observation of time-reversal symmetry violation through a
comparison of the probabilities for $\bar{K}^{0} \leftrightarrow
K^{0}$ oscillations as a function of the neutral-kaon proper
time.\cite{CPLEAR} The strangeness of the neutral kaon at the moment
of its creation, $t=0$, was tagged by observing the kaon charge in the
formation reaction $\bar{p}p \rightarrow K^{\pm}\pi^{\mp}(K^{0},
\bar{K}^{0})$ at rest, while the strangeness of the neutral kaon at
the time of its semileptonic decay, $t=\tau$, was tagged by the charge
of the final-state lepton.  The time-average decay-rate asymmetry,
measured over the interval $1\times\tau_{s} < \tau < 20\times
\tau_{s}$, is
\begin{equation}
	\left\langle \frac{\Gamma({\bar{K}^{0}}|_{0} \rightarrow 
	{e^{+}\pi^{-}\nu}|_{\tau}) - \Gamma({K^{0}}|_{0} \rightarrow 
	{e^{-}\pi^{+}\bar{\nu}}|_{\tau})}{\Gamma({\bar{K}^{0}}|_{0} \rightarrow 
	{e^{+}\pi^{-}\nu}|_{\tau}) + \Gamma({K^{0}}|_{0} \rightarrow 
	{e^{-}\pi^{+}\bar{\nu}}|_{\tau})}\right\rangle =
	(6.6 \pm 1.3_{\mathrm{stat}}\pm 1.0_{\mathrm{sys}}) \times 10^{-3}.
	\label{eq:CPLEAR}
\end{equation}
This asymmetry is a direct manifestation of
\textsf{T}-violation:\cite{rouge} If \textsf{CPT} is a good symmetry
in semileptonic decays and the $\Delta S=\Delta Q$ rule is exact, then
the observed asymmetry \eqn{eq:CPLEAR} is identical to
\begin{equation}
	\frac{{\mathcal P}(\bar{K}^{0} \rightarrow K^{0}) - {\mathcal P}(K^{0} \rightarrow \bar{K}^{0})}
	{{\mathcal P}(\bar{K}^{0} \rightarrow K^{0}) + {\mathcal P}(K^{0} \rightarrow 
	\bar{K}^{0})} ,
	\label{eq:probasym}
\end{equation}
where ${\mathcal P}$ is a probability for strangeness oscillation.  
The observed result is in good agreement with the theoretical 
expectation, $4\,\real{\varepsilon} = (6.63 \pm 0.06) \times 10^{-3}$.

At the same time, the KTeV Collaboration at Fermilab reported their
observation of a large \textsf{T}-odd effect in $1822 \pm 42$ examples
of the formerly rare $K_{L} \to \pi^{+}\pi^{-}e^{+}e^{-}$ decay
mode.\cite{kttime} An asymmetry of $(13.6 \pm 2.5 \pm 1.2)\%$ is seen
in the angular distribution between the $e^{+}e^{-}$ and
$\pi^{+}\pi^{-}$ decay planes, in the $K_{L}$ rest frame.  This is the
largest integrated \CP-violating effect yet observed, and is in
excellent agreement with theoretical predictions.\cite{sehgal,savage}

In Frascati we learned that the NA48 Collaboration has now identified
$458 \pm 22$ $K_{L} \to \pi^{+}\pi^{-}e^{+}e^{-}$ events and observed
the expected \textsf{T}-odd asymmetry of about $14\%$.\cite{na48rare}
In a \textsf{CPT}-invariant world, the observation of this
\textsf{T}-odd correlation would constitute direct evidence for
\textsf{T} noninvariance.  However, the right sort of \textsf{CPT}
violation could induce an asymmetry in the angular correlation without
the need to invoke \textsf{T} violation.\cite{bsT}

\subsection{New Limits on Charm Mixing}
Because the standard-model contributions to $D^{0}\hbox{ - 
}\bar{D}^{0}$ mixing and to \CP\ violation in $D$ decays are so 
minute, there are many opportunities to observe new 
physics.\cite{ikaros,gustavo}  The large number of fully 
reconstructed charmed mesons available in CLEO, the LEP experiments, 
and Fermilab fixed-target experiments make possible incisive 
searches, reviewed here by Jeff Appel.\cite{appel}  We can now 
contemplate experiments to reconstruct $10^{8}$ charms, and it is 
worth thinking about how to pursue those opportunities.  In 
particular, with Fermilab's 800-GeV fixed-target program at an 
end, we need to consider how to exploit dedicated $B$ experiments for 
charm.  A novel possibility, recently noticed, is that a 4-kg-year 
exposure at a neutrino factory could lead to a tagged sample of a 
million semileptonic $D$ decays.

\subsection{Other Rare Kaon Decays}
There is much work in progress on other rare---and formerly 
rare---kaon decays.  NA48 has used 74 examples of the double-Dalitz 
decay $K_{L} \to e^{+}e^{-}e^{+}e^{-}$ to verify the \CP-odd 
assignment of $K_{L}$.\cite{na48rare}  Let us note their plan for a 
future intense $K_{S}$ beam, with a goal of measuring the 
\CP-violation parameter $\eta_{000}$ in the $K_{S} \to 
\pi^{0}\pi^{0}\pi^{0}$ decay rate.  As for KTeV,\cite{marj} they have 
lowered the upper bounds
\begin{equation}
    B(K_{L} \to \pi^{0}e^{+}e^{-}) < 5.6 \times 10^{-10} \; ,
    \label{eq:klpiee}
\end{equation}
and
\begin{equation}
    B(K_{L} \to \pi^{0}\mu^{+}\mu^{-}) < 3.4 \times 10^{-10}
    \label{eq:klpimumu}
\end{equation}
to near the background levels.  They hope to triple their rare-decay 
data sets in current running.  For other evidence of progress on rare 
and radiative decays, see the talks by D'Ambrosio\cite{dam} and 
Kettell.\cite{kettell}

\subsection{Tests of \textsf{CPT} and Quantum 
Mechanics}
Finally, we heard brief reports on searches for violations of \textsf{CPT} 
invariance and failures of quantum mechanics in kaon 
decays.\cite{cplear}  We need to test these fundamental elements of 
physical theory, but it is difficult to know how to characterize 
deviations and what constitutes a viable theoretical 
framework.\cite{cpt,kostel}  I would caution that the link between 
string theory and observable deviations from quantum mechanics and 
\textsf{CPT} invariance is metaphorical at best.  It seems to me just 
slightly delusional to believe that in looking for violations of 
\textsf{CPT} and quantum mechanics one is testing the essentials of 
string theory in any direct way.

\section{Parting Thoughts}
The physics of \CP\ Violation and rare decays is in an exciting state:
We have new results to consider, we eagerly anticipate incisive new
information in the near future, and we have ambitious long-term
prospects.  \textsc{BaBar} and \textsc{belle}, which are running now, 
and CDF and D\O, which will run again in about a year, are ready to 
produce important results on $B$ mixing and decays.  The physics of 
neutrino masses, neutrino oscillations, and lepton flavor violation 
will soon join the physics of quark mixing as we seek to define and 
address the problem of identity.  Beginning next year, we expect 
great things from \textsc{kloe} on \CP\ violation and rare decays, and we 
wish \dafne\ a long and rich life!

\section{Acknowledgments}
I wish to compliment our hosts and organizers for the rich and varied 
program that brought many of us together for the first time, and for 
our warm reception in Frascati.  I particularly want to acknowledge 
the hospitality of Giorgio Capon, Franco Fabbri, Lia Pancheri, 
Juliette Lee-Franzini, and Paolo Franzini.  I thank the scientific
secretaries and members of the secretariat for their very efficient
work in scanning and photocopying transparencies from the
presentations, and for making them available so quickly on the
conference web site at \textsf{http://wwwsis.lnf.infn.it/dafne99/},
which greatly aided me in preparing this summary.  Fermilab is
operated by Universities Research Association Inc.  under Contract No. 
DE-AC02-76CH03000 with the United States Department of Energy.
\end{document}